\documentclass{segabs}
\usepackage{amsmath,amssymb,amsfonts}
\usepackage[colorlinks=true,urlcolor=blue,citecolor=black,linkcolor=black]{hyperref}

\usepackage{enumitem}

\usepackage[ruled,vlined,linesnumbered]{algorithm2e}

\usepackage{graphicx}
\usepackage{bm}
\usepackage{mathtools}

\usepackage{booktabs,xcolor,colortbl,amsmath,amssymb,array,caption,threeparttable,makecell}

\newcolumntype{L}[1]{>{\raggedright\arraybackslash}p{#1}}
\newcolumntype{C}[1]{>{\centering\arraybackslash}p{#1}}

\begin{document}

\title{Scalable Bayesian full waveform inversion via dual augmented Lagrangian SVGD}

\renewcommand{\thefootnote}{\fnsymbol{footnote}}

\author{Kamal Aghazade\footnotemark[1],
Ali Siahkoohi\footnotemark[2], and Ali Gholami\footnotemark[1] \\
  \footnotemark[1]Institute of Geophysics, Polish Academy of Sciences\\
 \footnotemark[2]Department of Computer Science, University of Central Florida
 }

\footer{Aghazade et al.}
\lefthead{Aghazade, Siahkoohi \& Gholami}
\righthead{Dual augmented Lagrangian SVGD for Bayesian FWI}

\maketitle

\begin{abstract}
Full waveform inversion is an ill-posed inverse problem whose solution
non-uniqueness---i.e., arising from band-limited, finite-aperture,
noisy data---calls for uncertainty quantification to avoid
overconfident geological interpretations. Bayesian inference addresses this need by characterizing the solution
as a posterior distribution rather than a single point estimate.
Sampling from this distribution, however, remains
computationally challenging: Markov chain Monte Carlo and non-amortized
variational inference require repeated wave equation solves, while
amortized variational inference approaches that avoid repeated solves
rely on training data that are inherently scarce in geoscience and face
unresolved generalization challenges in high dimensions. To address these
limitations, we integrate Stein variational gradient descent with the
alternating direction method of multipliers under a dual augmented
Lagrangian formulation. By fixing the wave operator at a background
model that is updated between frequency batches, it need only be
factorized once per particle per frequency, eliminating per-iteration
refactorization and reducing the total cost to that of a handful of
deterministic inversions while inheriting the favorable conditioning
of extended-space formulations. Applied to the Marmousi~II model, the
proposed method provides well-calibrated uncertainty estimates and
achieves inversion quality comparable to that of the standard
augmented Lagrangian SVGD at a fraction of the computational cost.
\end{abstract}
\vspace{-0.4cm}
\section{Introduction}

Full waveform inversion (FWI) estimates subsurface properties by
fitting seismic waveforms subject to the wave equation. The problem is
ill-posed---i.e., band-limited data, finite acquisition aperture, and
measurement noise render the solution non-unique---making uncertainty
quantification essential for reliable geological interpretation
\citep{Tarantola_1984_ISR,Virieux_2009_OFW}. Bayesian inference
addresses this need by
characterizing the solution as a posterior distribution over model
space, from which an ensemble of plausible models can be drawn to
compute posterior statistics such as means and standard deviations
\citep{Tarantola_1984_ISR,Stuart_2010_IPA}.

Generating samples from the posterior, however, remains a computational
bottleneck. Classical Markov chain Monte Carlo (MCMC) methods are
asymptotically exact but inherently sequential---i.e., each sample
depends on the previous one, ruling out parallelism across
samples---and mixing
degrades in high dimensions
\citep{Curtis_2001_PIS,Martin_2012_SNM,Zhao_2021_GMC,Zhang_2023_3DB}.
Amortized variational inference
\citep{Siahkoohi_2023_RAV,Baldassari_2024_CSD,Orozco_2025_AIR}
offers rapid posterior sampling by training a parametric conditional
distribution to approximate the posterior for previously unseen data,
but relies on a dataset of representative subsurface models, a
resource that is inherently scarce in geoscience applications. While
calibration techniques exist to mitigate generalization
errors---i.e., arising from limited training data and distribution
shifts at inference time
\citep{Siahkoohi_2023_RAV,Orozco_2025_AIR}---the extent to which they
can overcome overfitting when training data are limited in high
dimensions remains an open question
\citep{Baptista_2025_MRG,Siahkoohi_2026_OTR}.
Non-amortized parametric variational methods
\citep{Rizzuti_2020_PIU,Sun_2021_DPI,Zhao_2024_PSV,Zhao_2025_EBF}
avoid the need for training data but introduce a different tradeoff:
simple families such as mean-field Gaussians systematically
underestimate uncertainty \citep{Zhang_2023_3DB}, while expressive
alternatives such as normalizing flows
\citep{Rizzuti_2020_PIU,Yin_2025_WIS} require many trainable parameters
and face their own optimization challenges.

An alternative that avoids both the sequential nature of MCMC and the
parametric assumptions of variational inference is Stein variational
gradient descent \citep[SVGD;][]{Liu_2016_SVGD}. SVGD evolves an
ensemble of particles toward the posterior using only log-posterior
gradients, makes no distributional assumptions, and is embarrassingly
parallel across particles. Recent work has demonstrated its potential for Bayesian FWI
\citep{Zhang_2020_VFW,Zhang_2023_3DB,Corrales_2025_ASV}, yet the
computational cost remains prohibitive: each iteration requires
solving the Helmholtz equation for every particle, source, and
frequency via LU factorization, so that the total number of
factorizations scales with the product of particles, frequencies, and
iterations.

To address this challenge, we integrate SVGD with the alternating
direction method of multipliers under a dual augmented Lagrangian
formulation
\citep{Siahkoohi_2026_DSP,Aghazade_2025_FAF}. The augmented
Lagrangian relaxes the wave equation constraint and enforces it
progressively through multiplier updates, improving conditioning and
providing well-calibrated uncertainty estimates
\citep{Siahkoohi_2026_DSP}. By fixing the wave operator at a
background model \citep{Aghazade_2025_FAF}, it need only be factorized
once per particle per frequency, with all subsequent iterations
reducing to inexpensive forward--backward substitutions. For typical
particle counts, this places the cost of Bayesian uncertainty
quantification on par with that of a handful of deterministic
inversions.

We begin by formulating Bayesian FWI and its posterior distribution,
then derive the augmented Lagrangian sampling framework and the
SVGD-based algorithm. We next present the dual augmented Lagrangian
formulation that eliminates per-iteration factorization, and conclude
with numerical results on the Marmousi~II model.
\section{Bayesian full waveform inversion}

We consider frequency-domain FWI in the acoustic approximation. The
observed data $\bm{d}_i \in \mathbb{C}^{N_r}$ for the $i$-th source
experiment ($i = 1, \ldots, N_s$) are related to the subsurface
squared-slowness model $\bm{m} \in \mathbb{R}^N$ through
\begin{equation}
  \bm{d}_i = \bm{P}\bm{A}(\bm{m})^{-1}\bm{b}_i + \bm{n}_i,
  \label{eq:forward}
\end{equation}
where $\bm{A}(\bm{m}) = \omega^2 \operatorname{diag}(\bm{m}) + \Delta
\in \mathbb{C}^{N \times N}$ is the Helmholtz operator at angular
frequency $\omega$, $\bm{P} \in \mathbb{R}^{N_r \times N}$ is the
receiver sampling operator, $\bm{b}_i$ is the source term, and
$\bm{n}_i \sim \mathcal{N}(\bm{0}, \sigma^2 \bm{I})$. The forward
modeling operator $\bm{S}(\bm{m}) = \bm{P}\bm{A}(\bm{m})^{-1}$ maps
sources to predicted data at receiver locations. Because
$\bm{S}(\bm{m})$ is nonlinear in $\bm{m}$ and the data are
band-limited and aperture-limited, recovering $\bm{m}$ from
$\{\bm{d}_i\}$ is ill-posed: multiple models explain the observations
equally well. Bayesian inference addresses this non-uniqueness by
characterizing the solution as the posterior distribution, obtained via
Bayes' rule:
\begin{equation}
  p(\bm{m} \mid \bm{d})
  \;\propto\;
  \exp\!\Bigl(
    -\tfrac{1}{2\sigma^2}
    \textstyle\sum_{i=1}^{N_s}
    \|\bm{d}_i - \bm{S}(\bm{m})\bm{b}_i\|_2^2
  \Bigr)\,
  p(\bm{m}),
  \label{eq:posterior}
\end{equation}
where $p(\bm{m})$ is the prior distribution encoding prior knowledge
about the subsurface. An ensemble $\{\bm{m}^{(j)}\}_{j=1}^{N_p}$ can be drawn to
compute posterior statistics---i.e., means and standard
deviations---that quantify the range of plausible subsurface models
consistent with the observed data.
\vspace{-0.4cm}
\section{Sampling via the augmented Lagrangian}

Drawing samples from the posterior in equation~\eqref{eq:posterior}
requires solving $\bm{A}(\bm{m})^{-1}\bm{b}_i$ for every source at
every posterior evaluation, tightly coupling the model and wavefields
\citep{Fang_2018_UQE}. We relax this coupling by treating model
parameters and wavefields as independent variables and show that the
resulting augmented Lagrangian yields an evolving target distribution
amenable to particle-based sampling.

\subsection{Extended formulation}

As demonstrated by \citet{Fang_2018_UQE}, the tight coupling between
$\bm{m}$ and the wavefields in equation~\eqref{eq:posterior} produces
a posterior that inherits the full nonlinearity of the wave equation
solve, manifesting as spurious local modes in the negative
log-posterior landscape that severely hinder sampling. Treating
$\bm{m}$ and $\{\bm{u}_i\}$ as independent variables
\citep{Van_2013_MLM,Van_2016_PMF,Aghamiry_2019_IFM,Rizzuti_2021_DFW}
breaks this coupling:
\begin{equation}
  \min_{\bm{m},\,\{\bm{u}_i\}}
  \tfrac{1}{2\sigma^2}\textstyle\sum_{i=1}^{N_s}\|\bm{P}\bm{u}_i - \bm{d}_i\|_2^2
  \;\;\text{s.t.}\;\;
  \bm{A}(\bm{m})\bm{u}_i = \bm{b}_i.
  \label{eq:constrained}
\end{equation}
In this extended formulation, the wave equation is progressively
enforced rather than satisfied exactly at each iteration, significantly
reducing the ill-conditioning that plagues the reduced-space approach
\citep{Van_2013_MLM,Van_2016_PMF,Fang_2018_UQE,Aghamiry_2019_IFM,Operto_2023_ESS}.
While the
benefits of this relaxation for deterministic inversion are well
established, translating it into a posterior sampling framework requires
a probabilistic interpretation.

\subsection{Posterior distribution via the augmented Lagrangian}

To establish a probabilistic framework, we introduce Lagrange multipliers $\bm{v}_i \in \mathbb{C}^N$ and a
quadratic penalty, yielding the augmented Lagrangian
\citep{Aghamiry_2019_IFM,Gholami_2022_ESF,Gholami_2025_WLM}:
\begin{align}
  \mathcal{L}_\mu &= \tfrac{1}{2\sigma^2}
  \textstyle\sum_i \|\bm{P}\bm{u}_i - \bm{d}_i\|_2^2
  + \textstyle\sum_i \langle \bm{v}_i,\,
    \bm{A}(\bm{m})\bm{u}_i - \bm{b}_i \rangle
  \nonumber\\
  &\quad + \tfrac{\mu}{2}
  \textstyle\sum_i \|\bm{A}(\bm{m})\bm{u}_i - \bm{b}_i\|_2^2,
  \label{eq:AL}
\end{align}
where $\mu > 0$ is the penalty parameter. By introducing the scaled
multiplier $\bm{\varepsilon}_i = \bm{v}_i / \mu$ and eliminating
$\bm{v}_i$ at its saddle point, the augmented Lagrangian yields the
effective negative log-density over $\bm{m}$
\citep{Siahkoohi_2026_DSP}:
\begin{align}
  -\log p(\bm{m} \mid \bm{d}, \bm{\varepsilon})
  ={} & \tfrac{1}{2\sigma^2}
  \textstyle\sum_i \|\bm{P}\bm{u}_i - \bm{d}_i\|_2^2
  \nonumber\\
  +{} & \textstyle\sum_i \operatorname{Re}
    \langle \bm{\varepsilon}_i,\,
    \bm{A}(\bm{m})\bm{u}_i - \bm{b}_i \rangle
  \nonumber\\
  +{} & \tfrac{\mu}{2}
  \textstyle\sum_i \|\bm{A}(\bm{m})\bm{u}_i - \bm{b}_i\|_2^2
  - \log p(\bm{m}).
  \label{eq:neg_log}
\end{align}
In the above expression, the first term measures the data misfit, the
multiplier term steers the solution toward satisfying the wave
equation, and the penalty term quadratically penalizes current
constraint violations. For fixed multipliers $\bm{\varepsilon}_k$ at
iteration $k$, this defines an unnormalized distribution over
$\bm{m}$ that evolves as the multipliers are updated: in early
iterations, the distribution only coarsely approximates the true
posterior, yet as the constraint is increasingly enforced, it sharpens
and converges to $p(\bm{m} \mid \bm{d})$
\citep{Siahkoohi_2026_DSP}.

\subsection{Stein variational gradient descent}

To sample from this evolving target, we combine the alternating
direction method of multipliers (ADMM) with SVGD
\citep{Liu_2016_SVGD}, which evolves an ensemble of $N_p$
particles $\{\bm{m}^{(j)}\}_{j=1}^{N_p}$ toward the target
distribution. At each iteration the optimal displacement \citep{Liu_2016_SVGD} for particle
$j$ is 
\begin{align}
  \bm{\phi}^*(\bm{m}^{(j)})
  = \frac{1}{N_p} & \sum_{l=1}^{N_p}
  \Bigl[
    \underbrace{K(\bm{m}^{(l)}, \bm{m}^{(j)})
      \nabla_{\bm{m}^{(l)}} \log p(\bm{m}^{(l)} \mid \bm{d}, \bm{\varepsilon})
    }_{\text{driving force}}
  \nonumber\\
    &+
    \underbrace{\nabla_{\bm{m}^{(l)}} K(\bm{m}^{(l)}, \bm{m}^{(j)})
    }_{\text{repulsion}}
  \Bigr],
  \label{eq:svgd}
\end{align}
and each particle is updated as
$\bm{m}^{(j)} \leftarrow \bm{m}^{(j)} + \eta\,\bm{\phi}^*(\bm{m}^{(j)})$.
Here, $K$ is a radial basis function kernel with
bandwidth set by the median heuristic \citep{Liu_2016_SVGD}. The
driving force steers particles toward high-probability regions of
$p(\bm{m} \mid \bm{d}, \bm{\varepsilon})$, while the repulsive term
prevents collapse onto a single mode. To integrate SVGD with ADMM,
each particle carries its own auxiliary wavefields and scaled
multipliers. Defining
$\bm{S}^{(j)} = \bm{P}\bm{A}(\bm{m}^{(j)})^{-1}$ and
$\delta\bm{d}_i^{(j)} = \bm{d}_i - \bm{S}^{(j)}\bm{b}_i$, four
steps repeat at every iteration \citep{Siahkoohi_2026_DSP}:

\vspace{-1.em}
\begin{itemize}[leftmargin=1.em,itemsep=0pt]
\item \textbf{Step~1:} compute adjoint and forward wavefields for each particle with $\mu\sigma^2$ determined adaptively via the residual
whiteness principle \citep{Aghazade_2025_APP}.
\begin{align}
  \bm{\lambda}_i^{(j)} &=
  \bm{S}^{(j)\mathsf{H}}
  (\bm{S}^{(j)}\bm{S}^{(j)\mathsf{H}} + \mu\sigma^2\bm{I})^{-1}
  (\delta\bm{d}_i^{(j)} + \bm{S}^{(j)}\bm{\varepsilon}_i^{(j)}),
  \label{eq:adjoint}\\
  \bm{u}_i^{(j)} &=
  \bm{A}(\bm{m}^{(j)})^{-1}
  (\bm{b}_i + \bm{\lambda}_i^{(j)} - \bm{\varepsilon}_i^{(j)}).
  \label{eq:wavefield}
\end{align}
\item \textbf{Step~2:} compute the log-posterior gradient,
which has the familiar structure of a zero-lag cross-correlation between
forward and adjoint wavefields \citep{Plessix_2006_ARA}:
\begin{equation}
  \bm{g}^{(j)}
  =
  -\frac{1}{\omega^2}
  \frac{\sum_i \operatorname{Re}[(\bm{u}_i^{(j)})^* \circ \bm{\lambda}_i^{(j)}]}
       {\sum_i (\bm{u}_i^{(j)})^* \circ \bm{u}_i^{(j)}}
  + \nabla_{\bm{m}} \log p(\bm{m}^{(j)}),
  \label{eq:gradient}
\end{equation}
where the last term is the log-prior gradient, the
denominator provides source illumination compensation, $\circ$ denotes
elementwise multiplication, and ${}^*$ complex conjugation.
\item \textbf{Step~3:} update particles via
equation~\eqref{eq:svgd}.
\item \textbf{Step~4:} accumulate the wave equation
residual:
$\bm{\varepsilon}_i^{(j)} \leftarrow \bm{\varepsilon}_i^{(j)}
+ \bm{A}(\bm{m}^{(j)})\bm{u}_i^{(j)} - \bm{b}_i$.
\end{itemize}
\vspace{-1em}
Steps~1--2 are embarrassingly parallel across particles.
While \citet{Siahkoohi_2026_DSP} demonstrate that this framework
provides well-calibrated uncertainty estimates in high-dimensional
settings, Step~1 requires factorizing $\bm{A}(\bm{m}^{(j)})$ for
every particle at every iteration, yielding $N_p \times N_\omega
\times N_\text{iter}$ total LU factorizations, which is computationally expensive.
\vspace{-0.4cm}
\section{Scalable sampling via fixed operators}
The dual augmented Lagrangian formulation
\citep{Aghazade_2025_FAF} eliminates the per-iteration factorization
cost identified above. The key idea is to fix the wave operator at a background model
$\bm{m}_0^{(j)}$ for each particle, so that the Helmholtz operator
$\bm{A}_0^{(j)} \equiv \bm{A}(\bm{m}_0^{(j)})$ remains unchanged
throughout the inner iterations. The forward modeling operator
$\bm{S}_0^{(j)} = \bm{P}[\bm{A}_0^{(j)}]^{-1}$ and data residual
$\delta\bm{d}_{0,i}^{(j)} = \bm{d}_i - \bm{S}_0^{(j)}\bm{b}_i$ are
precomputed once per frequency, and the wavefield
equations~\eqref{eq:adjoint}--\eqref{eq:wavefield} become
\begin{align}
  \bm{\lambda}_i^{(j)} &=
  \bm{S}_0^{(j)\mathsf{H}}
  (\bm{S}_0^{(j)}\bm{S}_0^{(j)\mathsf{H}} + \mu\sigma^2\bm{I})^{-1}
  (\delta\bm{d}_{0,i}^{(j)} + \bm{S}_0^{(j)}\bm{\varepsilon}_i^{(j)}),
  \label{eq:adjoint_dual}\\
  \bm{u}_i^{(j)} &=
  [\bm{A}_0^{(j)}]^{-1}
  (\bm{b}_i + \bm{\lambda}_i^{(j)} - \bm{\varepsilon}_i^{(j)}).
  \label{eq:wavefield_dual}
\end{align}
Since $\bm{A}_0^{(j)}$ is fixed, its LU factors can be reused across
all inner iterations and source experiments via forward--backward
substitutions, at a fraction of the original factorization cost. The
multiplier update evaluates the constraint residual at the updated
particle:
\begin{equation}
  \bm{\varepsilon}_i^{(j)}
  \leftarrow
  \bm{\varepsilon}_i^{(j)}
  + \bm{A}(\bm{m}^{(j)} + \eta\,\bm{\phi}^*(\bm{m}^{(j)}))
    \bm{u}_i^{(j)} - \bm{b}_i.
  \label{eq:multiplier_dual}
\end{equation}
Note that evaluating $\bm{A}$ at the updated model requires no new
factorization as $\bm{A}(\bm{m})$ is linear in $\bm{m}$.
Fixing the operator is justified from two complementary perspectives.
From an optimization standpoint, the multiscale continuation strategy
ensures that the background model from the previous frequency band
already provides a good approximation, keeping the perturbations
$\delta\bm{m}$ small relative to $\bm{m}_0$; the multiplier updates in
equation~\eqref{eq:multiplier_dual} further compensate for any
mismatch by accumulating constraint residuals evaluated at the
updated model. From a sampling standpoint, fixing the operator is
transparent to SVGD: the gradient in
equation~\eqref{eq:gradient} is computed from the wavefields and
multipliers regardless of how those wavefields were obtained, so that
SVGD continues to transport particles toward the evolving target
without modification.
After the inner loop converges at a given frequency, the updated
particles serve as the background model for the next frequency in a
multiscale continuation strategy. The total number of LU factorizations
reduces from $N_p \times N_\omega \times N_\text{iter}$ to
$N_p \times N_\omega$, an $N_\text{iter}$-fold reduction. The
algorithm is summarized in Algorithm~\ref{alg:dualsvgd}.

\begin{algorithm}[!t]
\SetAlgoNlRelativeSize{0}
\setlength{\algomargin}{0pt}
\DontPrintSemicolon
\caption{Dual augmented Lagrangian SVGD}
\label{alg:dualsvgd}
\KwIn{Data $\{\bm{d}_i\}$, sources $\{\bm{b}_i\}$, prior
      $p(\bm{m})$, $N_p$, step size $\eta$}
\KwOut{Posterior samples
       $\{\bm{m}^{(j)}\}_{j=1}^{N_p}
       \approx p(\bm{m} \mid \bm{d})$}
\textbf{Init:} $\bm{m}_0^{(j)} \sim p(\bm{m})$,\;
  $\bm{\varepsilon}_i^{(j)} = \bm{0}$\\
\For{$\omega \in [\omega_{\min}, \omega_{\max}]$}{
  \For{$j = 1, \ldots, N_p$
       \textit{(embarrassingly parallel)}}{
    Factorize $\bm{A}_0^{(j)} = \bm{A}(\bm{m}_0^{(j)})$
      \hfill $\triangleright$ \textit{one LU per particle}\;
    Precompute $\bm{S}_0^{(j)}$, $\delta\bm{d}_{0,i}^{(j)}$\;
  }
  \For{$k = 1, \ldots, N_\text{inner}$}{
    \For{$j = 1, \ldots, N_p$
         \textit{(embarrassingly parallel)}}{
      Compute $\bm{\lambda}_i^{(j)}$
        via~\eqref{eq:adjoint_dual},\;
      $\bm{u}_i^{(j)}$
        via~\eqref{eq:wavefield_dual},\;
      $\bm{g}^{(j)}$
        via~\eqref{eq:gradient}\;
    }
    Update kernel bandwidth $h$ (median heuristic)\;
    \For{$j = 1, \ldots, N_p$
         \textit{(embarrassingly parallel)}}{
      $\bm{m}^{(j)} \leftarrow \bm{m}^{(j)} + \eta\,\bm{\phi}^{(j)}$
        via~\eqref{eq:svgd}\;
      Update $\bm{\varepsilon}_i^{(j)}$
        via~\eqref{eq:multiplier_dual}\;
    }
  }
  $\bm{m}_0^{(j)} \leftarrow \bm{m}^{(j)}$ for $j = 1, \ldots, N_p$
    \hfill $\triangleright$ \textit{update background}\;
}
\end{algorithm}

\begin{figure}[!t]
    \centering
    \includegraphics[width=0.95\columnwidth]{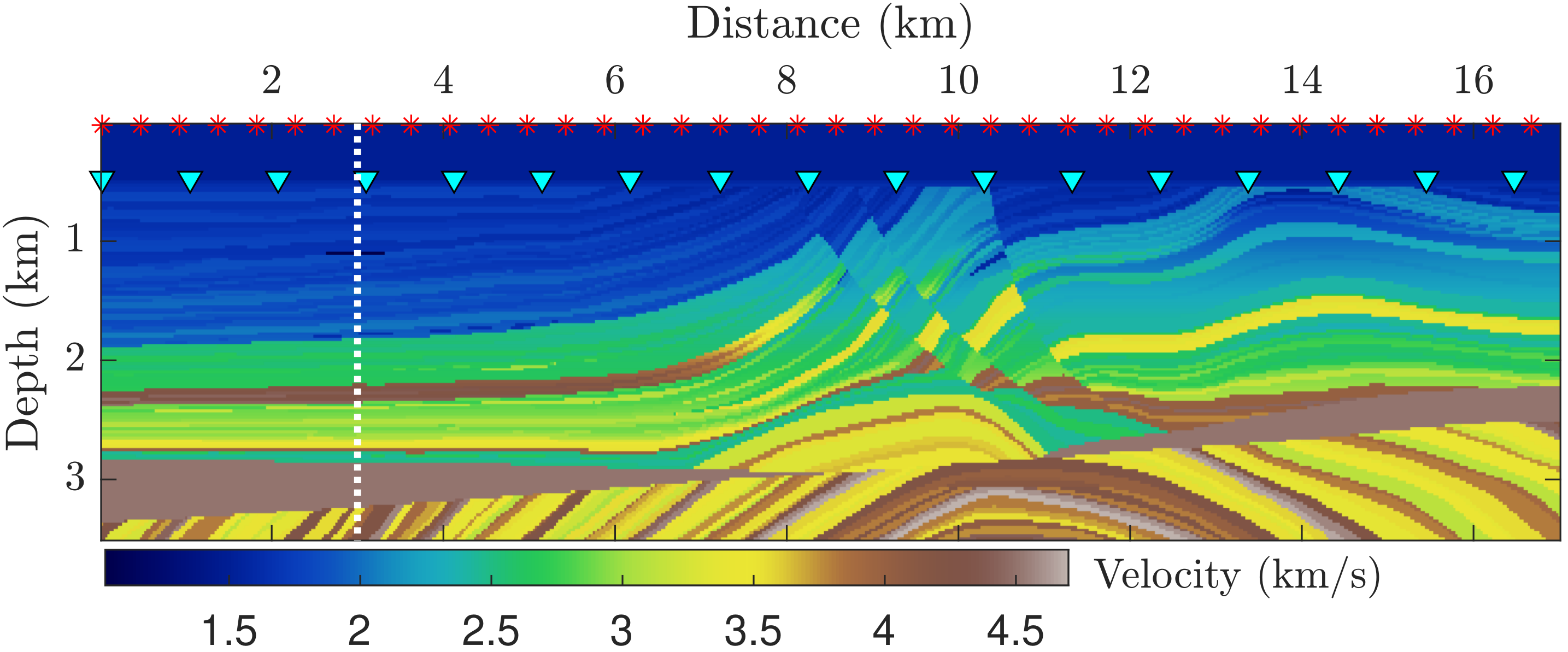}
    \caption{Marmousi~II P-wave velocity model with receivers
    (triangles) and sources (red stars).}
    \label{fig:model}
\end{figure}

\begin{table}[!t]
\caption{Per-particle runtime and complexity. AL: augmented
Lagrangian; Dual-AL: dual augmented Lagrangian.}
\label{tab:runtime}
\renewcommand{\arraystretch}{1.15}
\begin{tabular*}{\columnwidth}{@{\extracolsep{\fill}}lcccc@{}}
\hline
 & & AL & \multicolumn{2}{c}{Dual-AL} \\
\cline{3-3} \cline{4-5}
Step & Complexity & Time (s) & Complexity & Time (s) \\
\hline
LU of $\bm{A}$ & $\mathcal{O}(N^3)$ & $1.4$ & \multicolumn{2}{c}{precomputed once} \\
$\bm{S},\,\bm{S}^{\mathsf{H}}$ & $\mathcal{O}(2N_r N^2)$ & $6.3$ & \multicolumn{2}{c}{precomputed once} \\
$\bm{\lambda}$  & $\mathcal{O}(2N_s N^2)$ & $0.2$ & $\mathcal{O}(2N_s N^2)$ & $0.2$ \\
$\bm{u}$ & $\mathcal{O}(N_s N^2)$ & $1.5$ & $\mathcal{O}(N_s N^2)$ & $1.5$ \\
\hline
Total LU & \multicolumn{2}{c}{$N_p N_\omega N_{\mathrm{iter}}$} & \multicolumn{2}{c}{$N_p N_\omega$} \\
\hline
\end{tabular*}
\end{table}

\section{Numerical examples}

We apply the proposed method to the Marmousi~II velocity model
(Figure~\ref{fig:model}), a widely used test case featuring complex
geological structures including dipping layers, faults, and strong
lateral velocity variations. The model is 17.1~km wide and 3.5~km
deep, discretized on a 25~m grid ($685 \times 141$ grid points). A
total of $N_s = 34$ sources are uniformly distributed along the
surface and $N_r = 114$ receivers are uniformly spaced at 150~m
intervals. The prior distribution is a Gaussian random field with a
Mat\'{e}rn-type power spectrum and a linearly increasing background
velocity with depth \citep[see][for details]{Siahkoohi_2026_DSP}.
The ensemble consists of $N_p = 50$
particles evolved in parallel; a breakdown of per-particle runtime and
complexity is provided in Table~\ref{tab:runtime}. The inversion uses
a multiscale frequency-continuation strategy from 3~Hz to 12~Hz with a
0.5~Hz increment, implemented in three stages---(3--5.5)~Hz,
(3--8.5)~Hz, and (3--12)~Hz---with 10 inner iterations per frequency,
for a total of 370 iterations.

\begin{figure}[!t]
    \centering
    \includegraphics[width=\columnwidth]{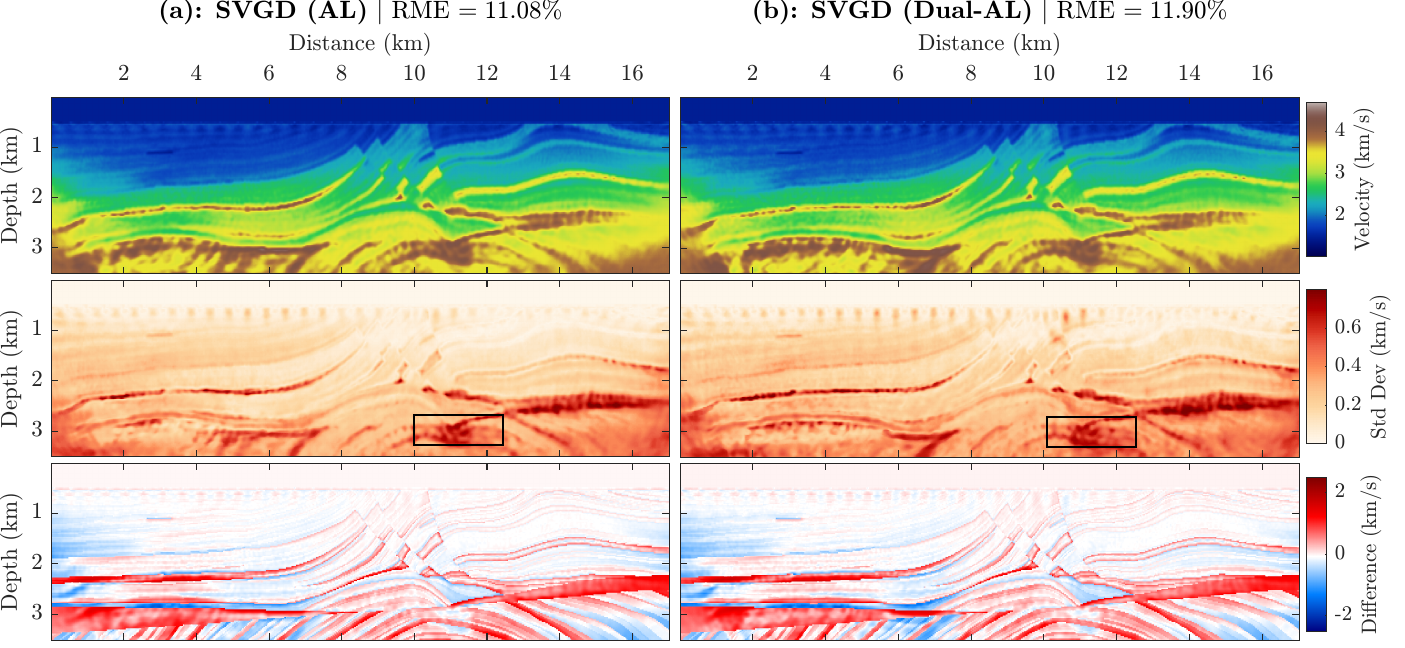}
    \caption{Augmented Lagrangian SVGD~(a) and dual augmented
    Lagrangian~(b) inversion results. Top: conditional mean with
    relative model error. Middle: pointwise standard deviation.
    Bottom: difference from the true velocity model.}
    \label{fig:UQ_res}
\end{figure}

Figure~\ref{fig:UQ_res} compares the conditional mean, posterior
standard deviation, and pointwise velocity difference for the
augmented Lagrangian and dual augmented Lagrangian SVGD formulations.
Both methods recover the main structural features of the Marmousi~II
model---including the dipping reflectors, fault zones, and velocity
inversions---with the augmented Lagrangian formulation achieving a
relative model error (RME) of 11.08\% compared to 11.90\% for the dual
augmented Lagrangian variant. The standard deviation panels reveal
coherent uncertainty patterns that correlate with areas of poor
illumination and strong lateral heterogeneity. The black rectangular
region highlighted in the standard-deviation panel corresponds to an
area of high posterior uncertainty, mainly due to limited angular
coverage.

\begin{figure}[!t]
    \centering
    \setlength{\tabcolsep}{3pt}
    \renewcommand{\arraystretch}{0}
    \begin{tabular}{@{}cc@{}}
    \includegraphics[width=0.34\columnwidth]{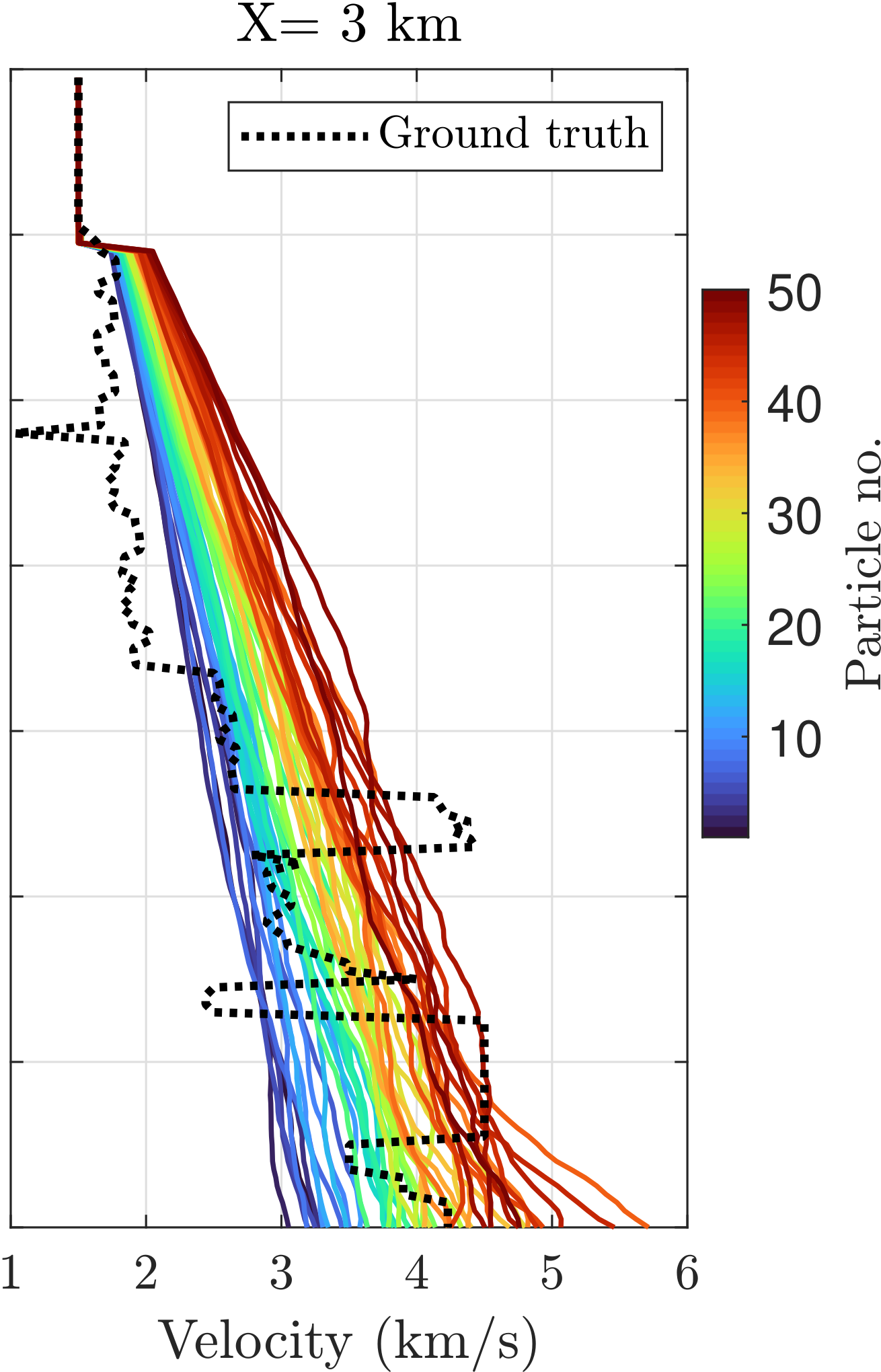} &
    \includegraphics[width=0.66\columnwidth]{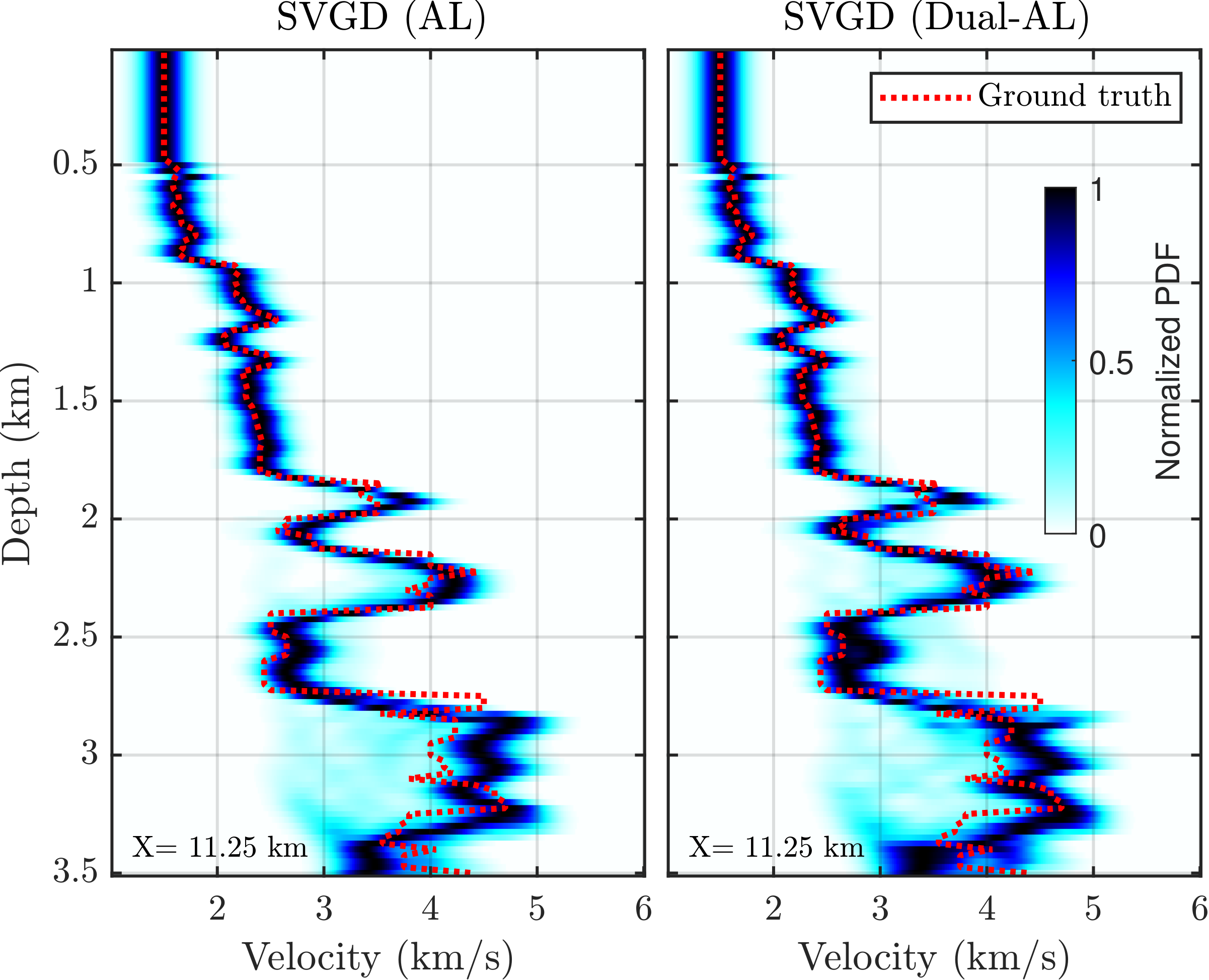} \\
    \end{tabular}
    \caption{(a) Initial particles (colored lines) and ground truth
    (black dotted line) at $X = 3$~km. (b,\,c) Normalized posterior
    probability density at $X = 11.25$~km for the augmented Lagrangian
    SVGD and its dual augmented Lagrangian counterpart. The red dotted
    line indicates the ground truth.}
    \label{fig:profiles}
\end{figure}

\begin{figure}[!t]
    \centering
    \includegraphics[width=1\columnwidth]{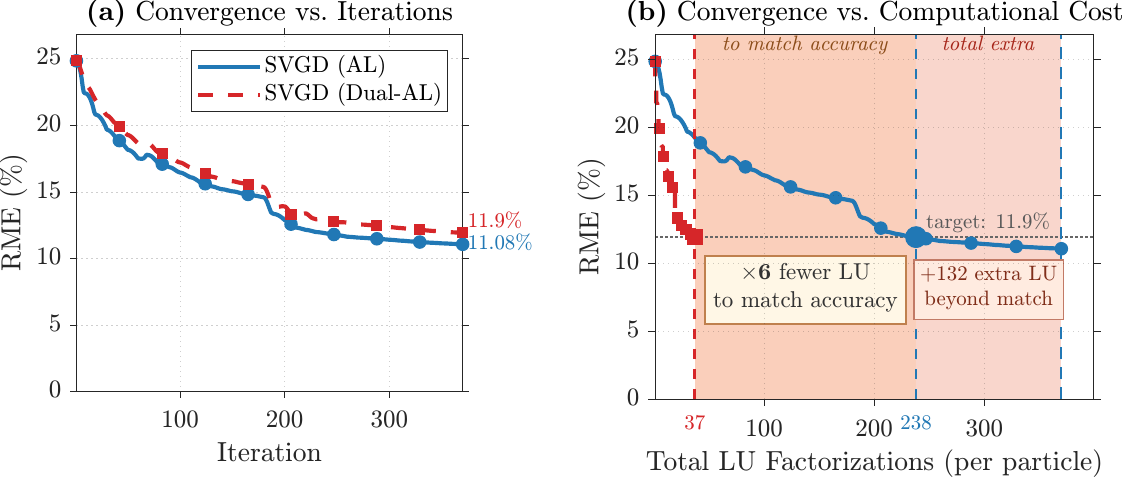}
    \caption{Relative model error versus (a) iteration count
    and (b) LU factorizations per particle.}
    \label{fig:RME}
\end{figure}

Figure~\ref{fig:profiles} presents the normalized posterior
probability density along a vertical profile at $X = 11.25$~km,
intersecting this zone. Despite successful recovery of mean
velocities, the posterior exhibits a broad spread of plausible
velocity values at greater depths, reflecting the inherent limitations
of surface-based seismic acquisition for resolving deep structure.
Figure~\ref{fig:RME} shows the convergence behavior in terms of RME.
Both formulations exhibit a consistent decrease in RME with iteration
count (panel~a), achieving comparable reconstruction accuracy. When
plotted against LU factorizations per particle (panel~b), the dual
augmented Lagrangian formulation requires approximately $37$
factorizations compared to $370$ for the augmented Lagrangian
SVGD---a tenfold reduction---while achieving comparable inversion
quality.
\section{Discussion and conclusions}

We have presented a dual augmented Lagrangian formulation for Bayesian
FWI that eliminates per-iteration refactorization of the wave operator,
reducing the cost of uncertainty quantification to that of a handful of
deterministic inversions. Despite fixing the wave operator, the
Marmousi~II results demonstrate comparable inversion quality and
uncertainty estimates to the standard augmented Lagrangian SVGD at a
fraction of the computational cost.
This is a consequence
of two mechanisms: the multiscale continuation strategy keeps the
perturbations small relative to the background model, and the
multiplier updates accumulate constraint residuals evaluated at the
updated model, progressively correcting for any mismatch. Natural directions for future work include extension to 3D---where
eliminating per-iteration factorizations yields even larger
savings---domain-specific kernel functions, and adaptive penalty
scheduling.
\vspace{-0.3cm}
\section{Acknowledgments}
{\tolerance=9999 Kamal Aghazade and Ali Gholami would like to thank the National
Science Center in Poland (grant no.\ 2022/46/E/ST10/00266).
Ali Siahkoohi acknowledges support from the Institute for
Artificial Intelligence at the University of Central Florida.\par}

\bibliographystyle{seg}
\bibliography{Mybiblio}

\end{document}